# Host Galaxy Dispersion Measure of Fast Radio Burst

Xinxin Wang[1], Ye-Zhao Yu[2, 3]

1. Valley Christian School, Dublin, 94568, United States of America;

2. Qiannan Normal University for Nationalities, Duyun, 558000, P. R. China; yuyezhao@foxmail.com

3. Qiannan Key Laboratory of Radio Astronomy, Duyun, 558000, P. R. China;

**Abstract**

Fast radio bursts are a class of transient radio sources that are generally thought to originate from extragalactic sources since their dispersion measure usually greatly exceeds the maximum dispersion measure that the Milky Way interstellar medium can provide. Host Galaxies of twenty-two fast radio bursts have already been identified. In this paper, the dispersion measure of these FRBs produced by the Milky Way interstellar medium and the intergalactic medium is obtained through known physical models to yield the host galaxy dispersion measure. It is found that the host galaxy dispersion measure increases with its redshift value, and that the host galaxy dispersion measure has different distribution between repeaters and non-repeaters. Further analysis suggests that there is no significant difference between the host galaxies of repeaters and non-repeaters, and the reason for the divergence of the host galaxy dispersion measures should be accounted for by the difference in their local environment.

## 1. INTRODUCTION

Fast Radio Burst (FRB) is a kind of radio transient with typical duration of several milliseconds and usually brighter than 1 Jy ms. Since the discovery of the first FRB (Lorimer et al. 2007), hundreds of FRB events have been detected (Petroff et al. 2016; CHIME/FRB Collaboration et al. 2021), which are divided into repeating and non-repeating bursts. Repeaters are FRBs that have been detected to occur twice or more (Spitler et al. 2016; CHIME/FRB Collaboration et al. 2019a; CHIME/FRB Collaboration et al. 2019b; Fonseca et al. 2020). Non-repeaters, on the other hand, are FRBs that have only been detected once so far. The difference in their origin is currently undetermined since we cannot conclude whether non-repeaters are also potential candidates for repeaters. Thus, although numerous of theoretical models have been proposed for FRB (Platts et al. 2019), its emission mechanism and physical origin still remain enigmatic. It has been suggested that non-repeating bursts are not unrepeating, but rather that their repeating bursts have not yet been detected by people, however, some studies argue that the probability that all FRBs are repeating bursts is extremely small (Palaniswamy et al. 2018; Caleb et al. 2019). It is commonly believed that repeaters and non-repeaters originate from different physical processes. Repeaters originate from non-catastrophic physical processes, such as giant pulses of young pulsars (Lyutikov, Burzawa & Popov 2016), giant flares of magnetar (Metzger, Margalit & Sironi 2019), accretion of white dwarfs by neutron stars (Gu et al. 2016), etc. Non-repeaters originate from catastrophic events such as the collapse of massive neutron stars into black holes (Falcke & Rezzolla 2014), binary neutron star mergers (Totani 2013), binary white dwarf mergers (Kashiyama, Ioka & Mészáros, 2013), neutron star-black hole mergers (Mingarelli, Levin & Lazio, 2015), etc.

A key parameter of FRBs is the dispersion measure (DM), which represents the integrated column density of free electrons between the observer and the signal source. Due to the presence of massive intergalactic and interstellar mediums in the space, the signals from FRB will be affected by these mediums as it travels through to the receiver, resulting in a delay in the reception of the low-frequency signal, which leads to the occurrence of dispersion. By comparing with the existing Milky Way electron distribution model, we can firmly indicate that the DM of FRB is greatly higher than the Milky Way DM contribution, indicating that FRB originates at an extragalactic distance.

As the FRB is considered to be extragalactic origin, the observed DM of FRB $DM_{obs}$ is modeled as

$$DM_{obs} = DM_{ISM} + DM_{halo} + DM_{IGM} + DM_{host}/(1+z) \quad (1)$$

where the $DM_{ISM}$, $DM_{\hbar alo}$, $DM_{IGM}$ and $DM_{\hbar ost}$ stand for the DM of interstellar medium(ISM), Milky Way halo, intergalactic medium(IGM), and host galaxy, respectively. $z$ is the redshift.

Among them, the DM values originated from Milky Way ($DM_{ISM}$) can be deduced from existing models. The three most widely used models are TC93 (Taylor, J. H. and Cordes, J. M. 1993), NE2001 (Cordes J. M., Lazio T. J. W., 2002), and YMW16 (Yao J. M., Manchester R. N., Wang N., 2017).

$DM_{\hbar alo}$ is the contribution of the Milky Way halo to the total dispersion measure, and the range of its value is considered to be 30 pc cm^-3 (Dolag et al.2015), 50-80 pc cm^-3 (Prochaska et al. 2019), or 43 pc cm^-3 based on the YT2020 Milky Way halo model (Yamasaki & Totani 2020).

Since FRBs are identified as extragalactic sources, we know that they should be located in their own host galaxies, finding the host galaxies of the FRBs therefore becomes an intriguing topic in the field of radio astronomy. However, there are only 22 FRBs discovered with known host galaxies (Heintz et al. 2020; Niu et al. 2022; Bhandari et al. 2022). The median value of DM contribution from the host galaxy is often considered to be 100 pc cm^-3 with a presented feature of log-normal distribution if the FRB is located in the galactic disk of a Milky Way-like host galaxy that can be represented by the NE2001 model and has an inclination angle of 60° (Thornton, D. et al. 2013; Xu, J., & Han, J. L. 2015). However, discrepancy can be caused by many factors. Not only do the angle of the observation, the type of the galaxy, the halo contribution, and the magnitude of the redshift generate deviations in the $DM_{\hbar ost}$ values, there is another important variable that is also relevant.

The DM contribution caused by the local environment of the FRB, also known as $DM_{source}$, describes the complexity of the environment the FRB is located at. The more intricate the local environment is, the greater the DM it will cause. For example, the FRB that is located in an extreme magneto-ionic environment or co-located with a compact, persistent radio source will have a higher $DM_{source}$ than the FRB located in a tranquil void region in the universe (Michilli, D. et al. 2018; Niu et al. 2022). In our discussion, $DM_{source}$ is included in $DM_{\hbar ost}$.

$DM_{IGM}$ is the dispersion measure distribution contribution of the intergalactic medium. Its value can be gained from subtracting $DM_{\hbar ost}$, $DM_{\hbar alo}$, and $DM_{ISM}$ from the total DM of FRB under certain circumstances. Thus, if we assume that the present-day electron numerical density of the intergalactic medium is uniformly distributed at $1.6 \times 10^{-7}$ cm^-3 (Katz, 2016), $DM_{IGM}$ can be used to estimate the distance of FRB from the observer or vice versa, which means that there are also some models that are able to calculate the $DM_{IGM}$ of the FRBs with known host galaxies and redshift based on the distance (Deng & Zhang 2014).

As more and more host galaxies of FRBs are observed, it provides a possibility to study the magnitude and distribution of the dispersion measure of FRB host galaxies. Bai 2022 tried to investigate $DM_{\hbar ost}$ with 13 FRBs of known host galaxies. They concluded that $DM_{\hbar ost}$ is nonlinearly related to the redshift value, and that there may be a linear relationship between $DM_{\hbar ost}$ of non-repeater and the galaxy metallicity. Lin H.-N. et al. 2022 studied 17 FRBs with known host galaxies and found no significant relationship between $DM_{\hbar ost}$ and host galaxy redshift, mass, star formation rate, and other parameters. However, in Lin H.-N. et al. 2022, the study did not investigate repeaters and non-repeaters separately.

In this paper, we collected samples of FRBs with identified host galaxies, and derived their host galaxy dispersion measures due to the influence of the Milky Way interstellar medium, Milky Way halo and intergalactic medium from existing models. By analyzing the dispersion measures of the host galaxies of repeaters and non-repeaters, we examined whether there is a difference between the galactic or local environments of these FRBs, thereby analyzing the difference in their physical origin.

## 2. DATA AND ANALYSIS

We collected 22 FRBs with known host galaxies and information about their host galaxies (Petroff at al. 2016; Heintz et al. 2020; Niu et al. 2022; Bhandari et al. 2023; Connor et al. 2023), including the observed dispersion measure ($DM_{obs}$) and the redshift value ($z$). We estimated the dispersion measure contribution from the Milky Way interstellar medium ($DM_{ISM}$), based on the YMW16 galactic electron density model (Yao, Manchester & Wang, 2017), and the dispersion measure contribution from the Galactic Halo ($DM_{\hbar alo}$) based on the YT2020 model (Yamasaki & Totani, 2020), as listed in Table 1. In the “Rep” column of the table, the value of 0 denotes that the FRB is a non-repeater, and the value of 1 denotes that it is a repeater; $z$ is the host galaxy redshift value; offset is the projected distance of the FRB from the host galaxy galactic center; R is the effective radius of the host galaxy (Peng et al. 2010); SFR is the host galaxy star formation rate; M is the mass of the host galaxy.

Since the host galaxies are known, the distance to the FRBs can be considered as the distance to the host galaxies, and thus the dispersion measure contribution of the intergalactic medium ($DM_{IGM}$) can be estimated according to Eq. 2 (Deng & Zhang,

Table 1. Parameters of FRBs with known host galaxies.

| Name | $DM_{obs}$ pc $cm^{-3}$ | $DM_{ISM}$ pc $cm^{-3}$ | $DM_{halo}$ pc $cm^{-3}$ | Rep | z | offset kpc | R kpc | SFR $M_\odot$ $yr^{-1}$ | M $M_\odot$ |
|---|---|---|---|---|---|---|---|---|---|
| 20121102A | 557 | 287.0788 | 40.91401 | 1 | 0.1927 | 0.75 | 2.05 | 0.15 | 143000000 |
| 20180301A | 536 | 253.9594 | 38.05757 | 1 | 0.3305 | 10.8 | 5.8 | 1.93 | 2300000000 |
| 20180916B | 348.8 | 324.8824 | 43.07673 | 1 | 0.0337 | 5.46 | 3.57 | 0.06 | 2150000000 |
| 20180924B | 362.16 | 27.6485 | 45.54362 | 0 | 0.3214 | 3.37 | 2.75 | 0.88 | 13200000000 |
| 20181030A | 103.5 | 33.05255 | 31.57422 | 1 | 0.0039 | 0 | 2.6 | 0.36 | 5800000000 |
| 20181112A | 589 | 29.0287 | 45.0438 | 0 | 0.4755 | 1.69 | 7.19 | 0.37 | 3980000000 |
| 20190102C | 364.545 | 43.28 | 46.98819 | 0 | 0.2913 | 2.26 | 5 | 0.86 | 3390000000 |
| 20190520B | 1204.7 | 50.24369 | 69.1571 | 1 | 0.241 | * | * | 0.41 | 600000000 |
| 20190523A | 760.8 | 29.87962 | 32.39296 | 0 | 0.66 | 27.2 | 3.28 | 0.09 | 61200000000 |
| 20190608B | 340.05 | 26.62266 | 38.89012 | 0 | 0.1178 | 6.52 | 7.37 | 0.69 | 11600000000 |
| 20190611B | 332.63 | 43.67048 | 47.21411 | 0 | 0.3778 | 11.7 | 2.15 | 0.27 | 750000000 |
| 20190711A | 592.6 | 42.61163 | 46.20018 | 1 | 0.5217 | 3.17 | 2.94 | 0.42 | 810000000 |
| 20190714A | 504.13 | 31.15956 | 36.43899 | 0 | 0.2365 | 2.7 | 3.94 | 0.65 | 14200000000 |
| 20191001A | 507.9 | 31.08082 | 46.71377 | 0 | 0.234 | 11.1 | 5.55 | 8.06 | 46400000000 |
| 20191228A | 297.5 | 19.92478 | 36.87967 | 0 | 0.2432 | 5.7 | 1.78 | 0.03 | 5400000000 |
| 20200120E | 87.82 | 32.50943 | 31.12772 | 1 | 0.00014 | 20 | 3.5 | 0.6 | 72000000000 |
| 20200430A | 380.25 | 26.07647 | 42.24145 | 0 | 0.1608 | 1.7 | 1.64 | 0.26 | 2100000000 |
| 20200906A | 577.8 | 37.86544 | 30.16039 | 0 | 0.3688 | 5.9 | 7.58 | 0.48 | 13300000000 |
| 20201124A | 413.52 | 196.6219 | 36.23963 | 1 | 0.0979 | 1.3 | * | 2.12 | 16000000000 |
| 20210117A | 728.95 | 21.43 | 35.37 | 0 | 0.214 | 2.8 | * | 0.014 | 363078054 |
| 20220509G | 269.53 | 52.07 | 37.7 | 0 | 0.0894 | * | * | * | * |
| 20220914A | 631.29 | 51.11 | 36.95 | 0 | 0.1139 | * | * | * | * |

2014; Zhang, 2018).

$$DM_{IGM} = \frac{3cH_0\Omega_b f_{IGM}}{8\pi G m_p} \times \int_0^z \frac{\left[\frac{3}{4}y_1\chi_{e,H(z)}+\frac{1}{8}y_2\chi_{e,He(z)}\right](1+z)dz}{[\Omega_m(1+z)^3+\Omega_\Lambda]^{1/2}} \quad (2)$$

where $c$ is the speed of light in vacuum, $G$ is the universal gravitational constant, $m_p$ is the mass of proton, $H_0 = 67.36\ \mathrm{km \cdot s^{-1} \cdot Mpc^{-1}}$ is the Hubble constant, $\Omega_b = 0.0493$ is the percentage of baryonic matter mass, and $f_{IGM} = 0.83$ is the percentage of baryonic matter mass in the intergalactic medium. $\frac{3}{4}y_1\chi_{e,H(z)} + \frac{1}{8}y_2\chi_{e,He(z)}$ represents the degree of ionization of the intergalactic medium. Assuming that the proportion of hydrogen in the intergalactic medium is 3/4 and the proportion of helium is 1/4, the coefficient $y_1 \sim y_2 \sim 1$. Assuming complete ionization of the intergalactic medium, the ionization degree of hydrogen $\chi_{e,H(z)} \sim 1$ and that of helium $\chi_{e,He(z)} \sim 1$. $\Omega_m = 0.315$ is the matter density parameter. $\Omega_\Lambda = 0.6911$ is the dark energy density parameter.

According to Eq. (1), we learn that the dispersion measure contribution from host galaxy should be:

$$DM_{host} = (1+z)(DM_{obs} - DM_{ISM} - DM_{halo} - DM_{IGM}) \quad (3)$$

Upon computation of the $DM_{host}$, we proceeded to construct a scatter plot denoted as Figure 1, depicting the $DM_{host}$ against the redshift values $z$ in an attempt to find any potential correlations between the two. Notably, within our data set, we identified three FRBs with $DM_{host}$ below 0 due to inherent uncertainties in estimating $DM_{ISM}$, $DM_{halo}$, and $DM_{IGM}$. As can be seen from Fig. 1, there is a tendency for the $DM_{host}$ to increase with the redshift value $z$, and the relationship can be demonstrated by Eq. 4 (Zhang G. Q. et al. 2020):

$$DM_{host} = A(1+z)^\alpha \quad (4)$$

in which we derive by fitting that $A = 188.84 \pm 99.39$, and $\alpha = 0.70 \pm 1.96$.

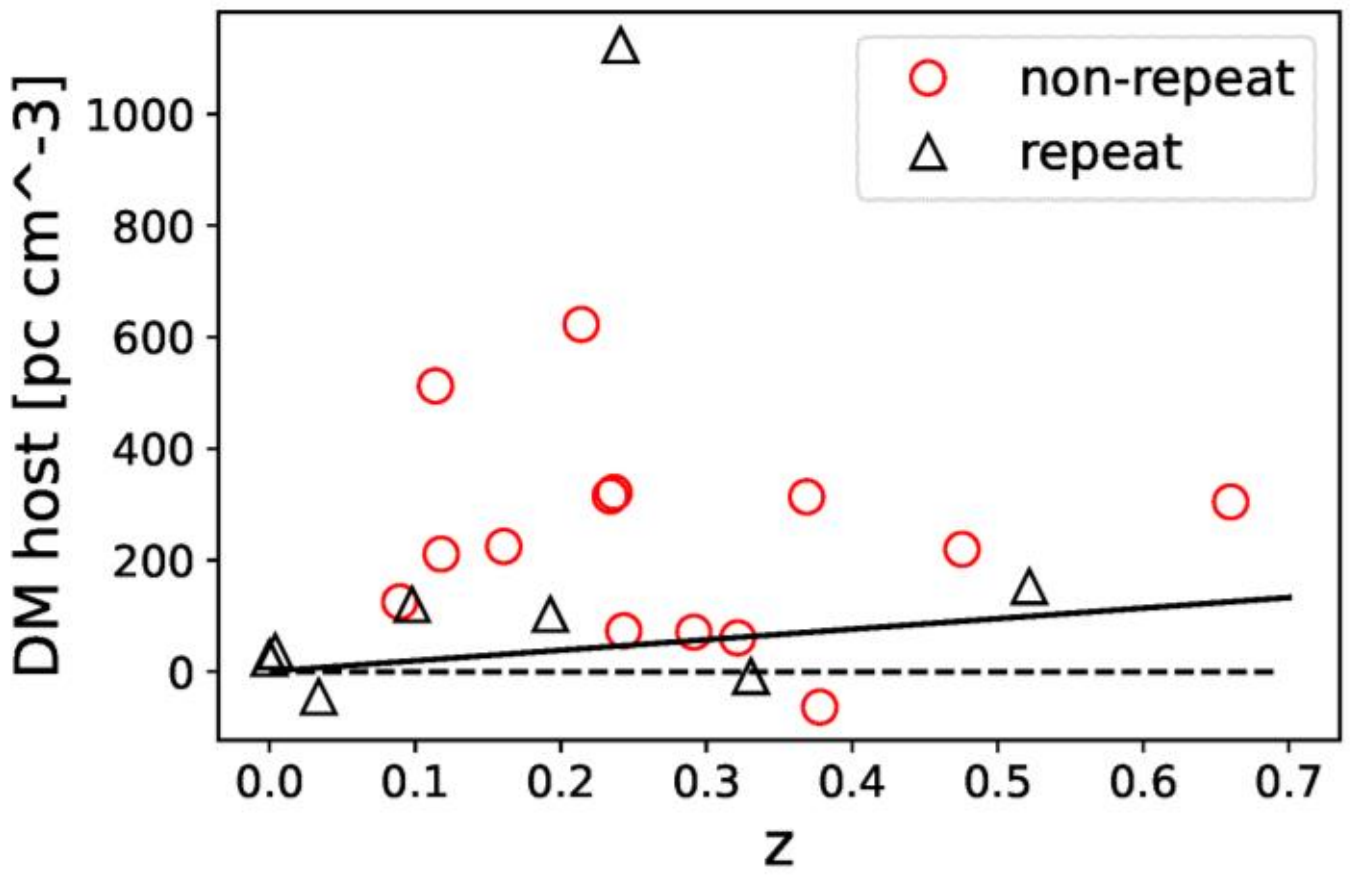

Figure 1. Plot of the $DM_{host}$ against redshift value $z$. The red circles are non-repeaters. The black triangles are repeaters. The black solid line is the result of the linear fitting of all samples. The black dashed line is the position where the dispersion measure value is 0.

We also noticed from Fig. 1 that the distribution of $DM_{host}$ of repeaters and non-repeaters is different, and repeaters exhibit generally lower $DM_{host}$ compared to non-repeaters. Among them, FRB 20190520B, discovered by the FAST telescope, is a repeater with a significantly higher $DM_{host}$ than other repeaters.

To comprehensively investigate this discrepancy, we conducted an analysis of $DM_{host}$ (samples no less than 0) for repeaters and non-repeaters and plotted histogram for them. The distribution of $DM_{host}$ can be described by a log-normal distribution (Zhang G. Q. et al. 2020), so the $DM_{host}$ in the histogram is taken as logarithm with a base of 10, and the result is depicted in Fig. 2. In addition, we fit the histogram with a Gaussian function. The fitted histogram shows that the mean value of non-repeater's $DM_{host}$ is $167.27^{+96.95}_{-61.37}$ pc cm^3 and the repeater's $DM_{host}$ contribution is $111.89^{+315.27}_{-82.58}$ pc cm^3.

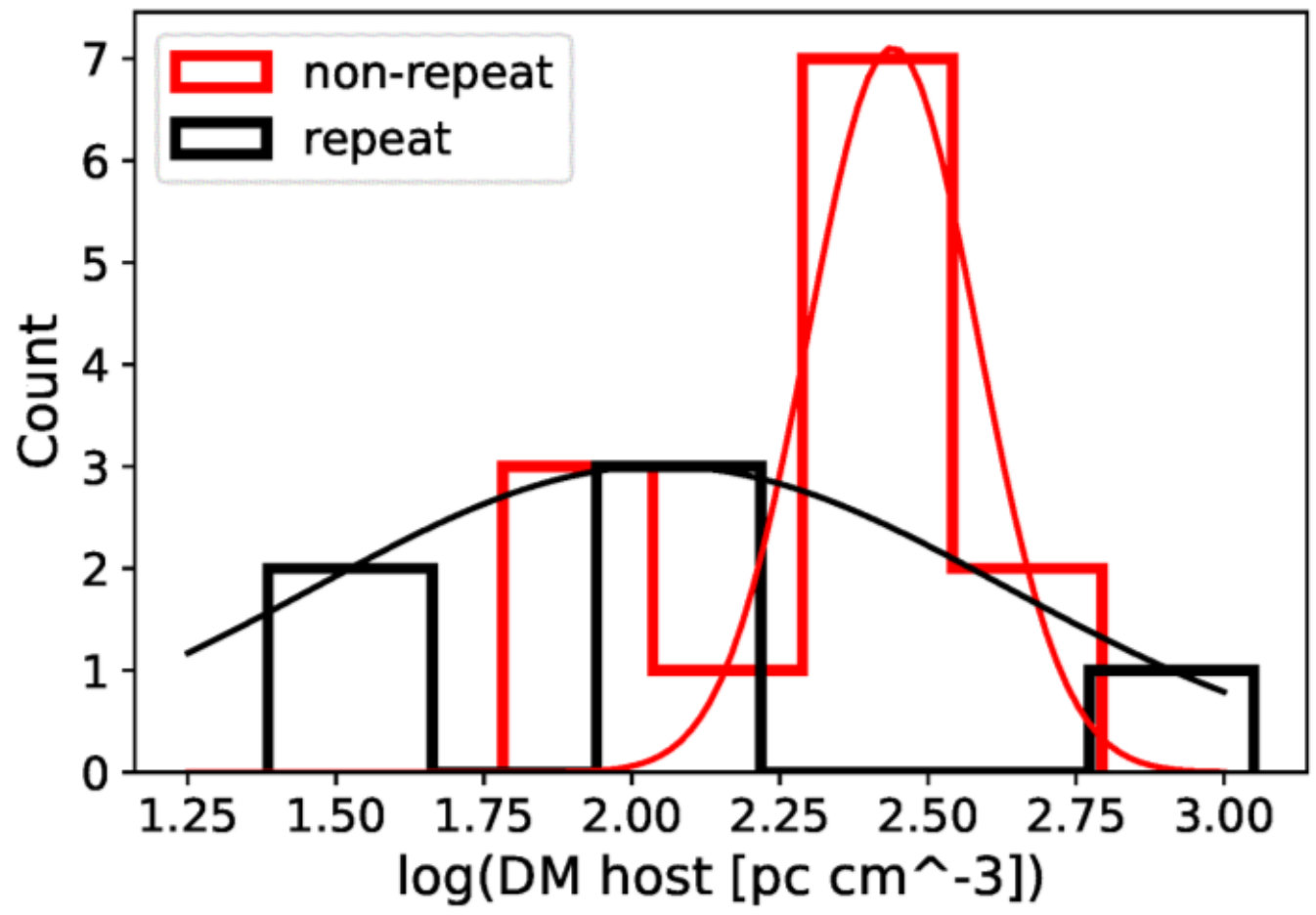

Figure 2. Histogram of the $DM_{host}$. The red line is the histogram of non-repeaters. The red curve is the result of its corresponding gaussian function fit. The black line is the histogram of repeaters. The black curve is the result of its corresponding Gaussian fit.

## 3. DISCUSSION AND CONCLUSIONS

As seen in Fig. 1, $DM_{host}$ of FRB increases with increasing redshift. This trend is caused by the cosmological evolution and is consistent with the trend obtained by previous simulations(Jaroszynski, M. 2020; Zhang G. Q. et al. 2020).

From Fig. 2, we notice a clear difference in the distribution of $DM_{host}$ between repeating and non-repeating bursts, and $DM_{host}$ comes mainly from two components: one is the contribution of the host galaxy interstellar medium, and the other is the contribution of the surrounding substance of the FRB. It should be the differences between these two components or one of them that lead to the difference in $DM_{host}$ distribution of repeaters and non-repeaters. Generally, we can simply assume that the closer the position is to the galactic center, the larger the DM in the host galaxy of the FRB is. For this reason we first analyzed the discrepancy between the locations of repeaters and non-repeaters in host galaxies to investigate whether it is the difference in $DM_{host}$ that causes the difference in their distribution. To make a better comparison among samples, we begin by defining the relative deviation of FRB from its host galaxy galactic center:

$$R_{offset} = offset/R \quad (5)$$

where the $offset$ is the projected distance of the FRB from the center of the host galaxy; R is the effective radius of the host galaxy. Further, the relationship between $R_{offset}$ and $DM_{host}$ was analyzed. As shown in Fig. 3, $DM_{host}$ generally exhibits a decreasing tendency with increasing $R_{offset}$, which is in line with our general expectation. There is no significant difference in the distribution of $R_{offset}$ between repeaters and non-repeaters, but it can be noted that the non-repeaters near the center of the host galaxy have a larger amount of $DM_{host}$. This may implicate a difference in the nature of the host galaxy between repeaters and non-repeaters.

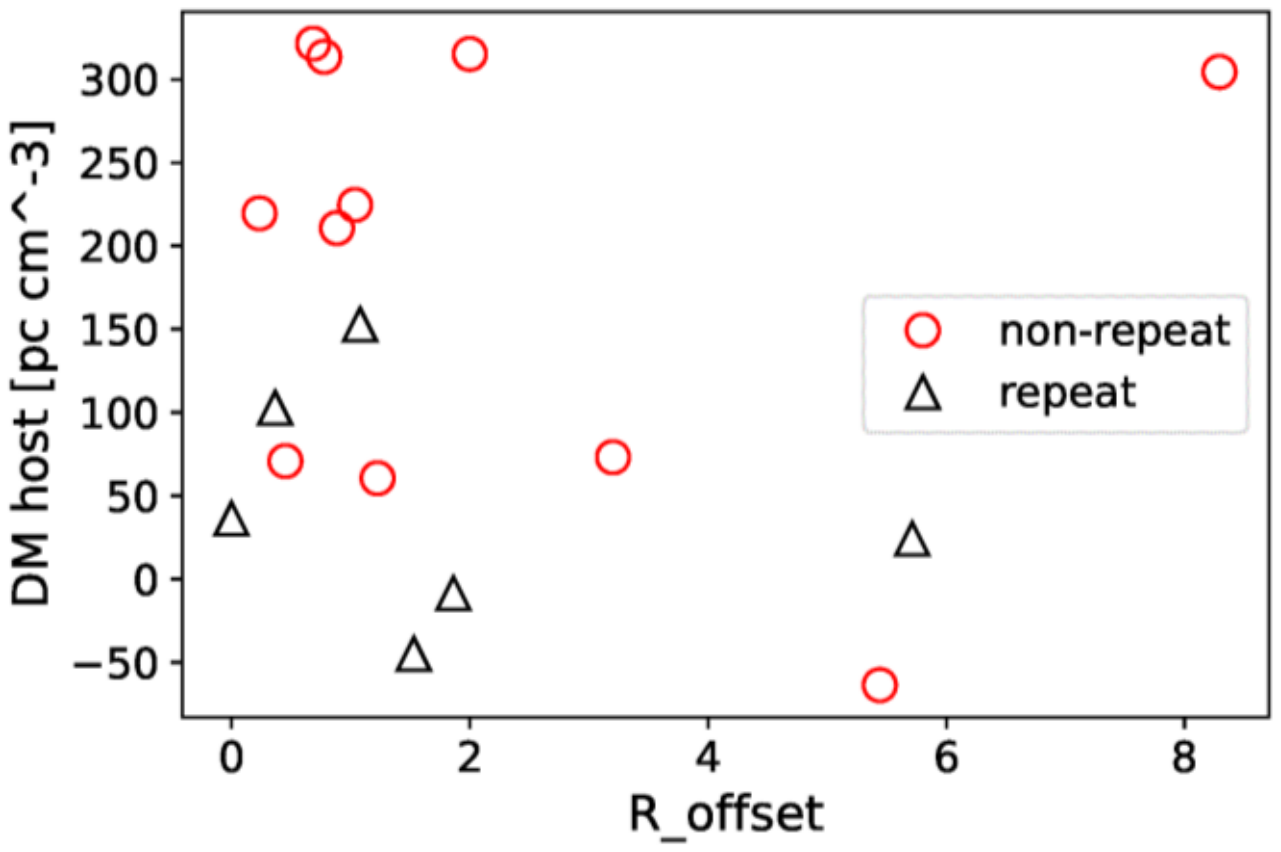

Figure 3. Plot of the relative offset value of FRB from the galactic center of the host galaxy versus the $DM_{host}$ contribution.

We analyzed the relationship between the star-formation rate and $DM_{host}$ contribution of repeaters and non-repeaters. As illustrated in Fig. 4, the host galaxy star-formation rate of non-repeaters has a large distribution, ranging from 0.01 to 10 $M_{\odot} yr^{-1}$, while the host galaxy star formation rate of repeaters is relatively concentrated between 0.1 to 3 $M_{\odot} yr^{-1}$.

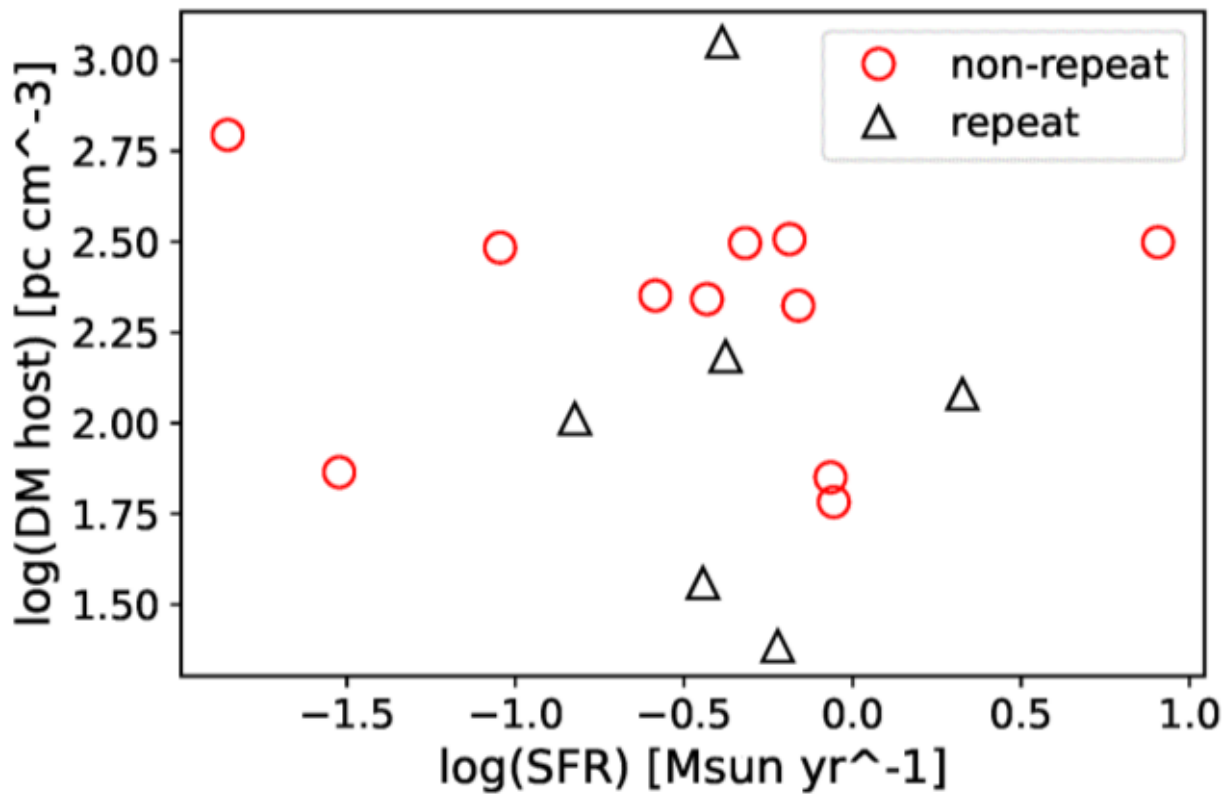

Figure 4. Plot of FRB host galaxy SFR versus DM host contribution.

We further investigate whether these galaxies differ by analyzing the relationship between the mass of the host galaxy and the star formation rate. As displayed in Fig. 5, a plot of the relationship of the two variables mentioned above between repeaters and non-repeaters is shown, in which the black dashed line is an approximate boundary, described by Eq. 6 (Chen et al. 2016; Jin et al. 2016).

$$\log(SFR) = 0.86 \times \log(M) - 9.29. \quad (6)$$

Above this boundary are the star-forming galaxies. A few host galaxies lie below the boundary are belong to green valley galaxies. In Fig. 5, there is no significant difference in the distribution of the host galaxies of repeaters and non-repeaters. Thus, we suggest that the variation in the $DM_{host}$ distribution between repeaters and non-repeaters is not driven by the difference between the host galaxies, but by the difference in the local environment of the FRB.

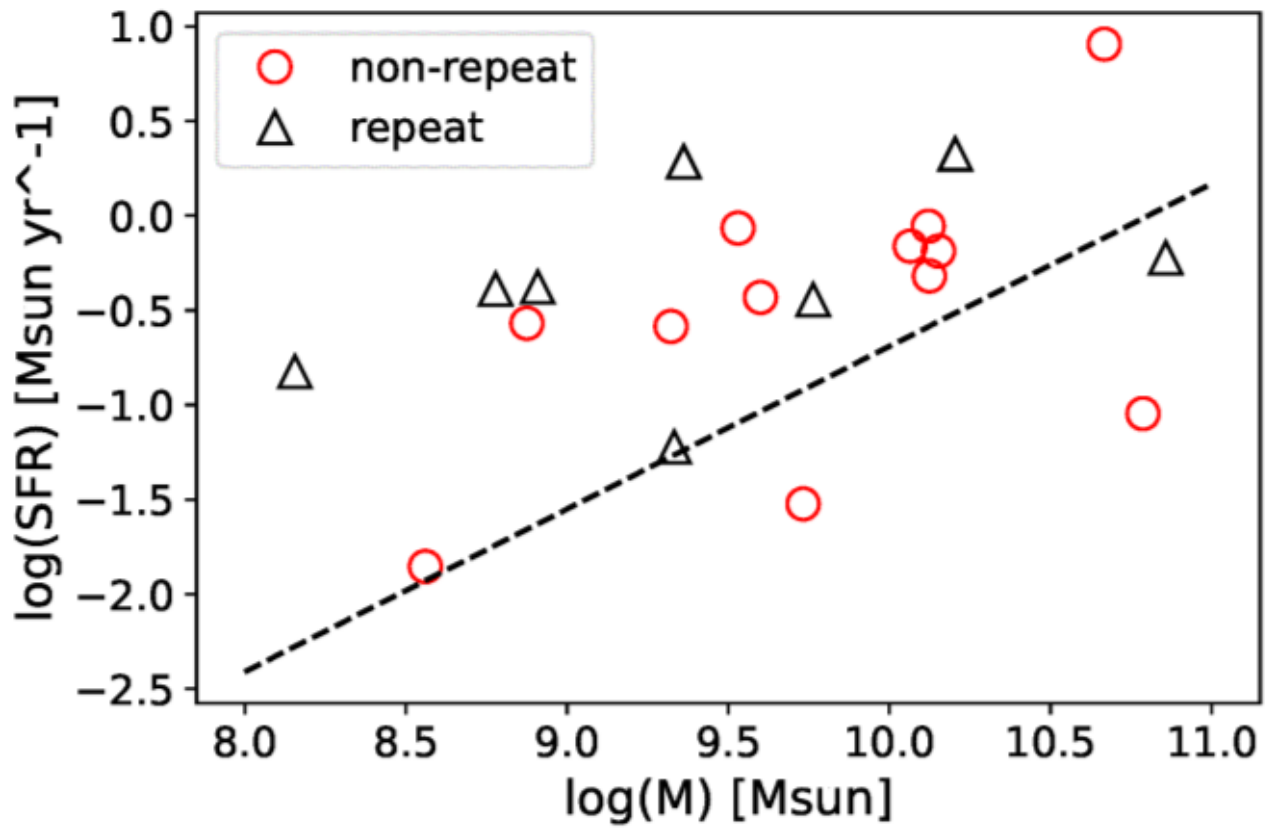

Figure 5. Plot of FRB host galaxy mass (M) versus star-formation rate (SFR). The black dashed line is the approximate boundary. Above this boundary are Star-forming galaxies and below it are Green Valley galaxies.

We have estimated $DM_{host}$ of FRBs with known host galaxies by available physical models. The analysis finds that (1) $DM_{host}$ increases with redshift $z$, and this relation can be described by $DM_{host} = 188.84 \pm 99.39 \times (1+z)^{0.7\pm1.96}$; (2) the $DM_{host}$ distribution of repeaters has a smaller mean value and a larger distribution compared with that of non-repeaters. The mean value of $DM_{host}$ of repeaters is $111.89^{+315.27}_{-82.58}$ pc cm^3, and that of non-repeaters is $275.81^{+107.50}_{-77.35}$ pc cm^3. Further analysis reveals that: (1) there is no significant difference between the distribution of repeaters and non-repeaters relative to the galactic centers, but the $DM_{host}$ of non-repeaters near the galactic center is significantly larger; (2) the star-formation rate of the host galaxy of non-repeaters is more widespread, ranging from 0.01 to 10 M⊙ yr$^{-1}$, while the star formation rate of the host galaxy of repeaters is relatively concentrated from 0.1 to 3 M⊙yr$^{-1}$; (3) there is no significant difference between the host galaxy of repeaters and non-repeaters in the distribution of log(M)-log(SFR) diagram. Most of the host galaxies belong to star-forming galaxies, and a few belong to green valley galaxies. In summary, we suggest that there is a difference between the local environment of repeaters and non-repeaters, with more free electrons in the local environment of non-repeaters, leading to a higher $DM_{host}$. This finding may affect the estimation of the circumstellar magnetic field of repeaters and non-repeaters. If $DM_{host}$ of repeaters is indeed larger than that of non-repeaters due to the influence of local matter, as we have obtained, this could to some extent restrict the physical models of the origin of repeaters and non-repeaters. For example, in this case, the origin of repeaters is less likely to be giant pulse of young pulsars since young pulsars are more likely to have a larger $DM_{host}$ due to the association with supernova remnants.

**ACKNOWLEDGEMENTS：** This work is supported by the Science and Technology Program of Guangzhou under grant number 202102010466.

**REFERENCES：**